\documentstyle[a4,11pt,epsfig,twoside]{article}
\def\ggg{$\Gamma_{\gamma \gamma}$}

\def\pgg{$\phi_{\gamma \gamma}$}

\def\epem{$e^+ e^-$}

\def\z0{Z}

\begin{document}

\title{TWO-PHOTON WIDTH AND  PHASE MEASUREMENT FROM  
       THE SM HIGGS DECAYS INTO $WW$ AND $ZZ$   
            AT THE PHOTON COLLIDER} 

\author{P. NIE\.ZURAWSKI$^1$,
        A.F. \.ZARNECKI$^1$
        and  M. KRAWCZYK$^{2,3}$
\\
\\
    $^1$ {\it Institute of Experimental Physics, Warsaw University, Poland}
\\
    $^2$ {\it Theory Division, CERN, Switzerland}
\\
    $^3$ {\it Institute of Theoretical Physics, Warsaw University, Poland} 
}

\date{}

\maketitle

\begin{abstract}
Production of the Standard Model Higgs-boson at the photon collider at
TESLA is studied for the Higgs-boson masses above 170 GeV. 
By reconstructing $W^+ W^-$ and $\z0  \z0  $ final states,
not only the $h \rightarrow \gamma \gamma$ partial width 
but also the phase of the scattering amplitude can be measured.
 This opens a new window onto the precise
determination of the Higgs-boson couplings.
\end{abstract}


A photon collider has been  proposed as a natural
extension of the \epem\ linear collider TESLA \cite{tdr_pc}.
It is the ideal place to study the properties of the Higgs-boson
and the electroweak symmetry breaking (EWSB).
This contribution summarizes results of \cite{wwzz_paper},
where the feasibility of measuring  Higgs-boson production
in $W^+ W^-$ and $\z0 \z0$ decay channels has been studied
for Higgs-boson mass above 170~GeV.


We study the signal, ie. the Higgs-boson decays into vector bosons,
and the background from direct vector-bosons production.
For the $\z0  \z0 $ final state the direct, ie. non-resonant 
$\gamma \gamma \rightarrow \z0 \z0$ process is possible 
at the loop level only, while the non-resonant $W^+W^-$ pair 
production is a tree-level process, and is expected to be large.
Also an interference between the signal of $W^+W^-$ production 
via the Higgs resonance and the background from direct production
can be large.
The measurement of the interference contribution
allows us to access an information about the phase of
the $h \rightarrow \gamma \gamma $ amplitude, \pgg.
For the Higgs-boson masses around 350 GeV, we found that
the amplitude phase \pgg\ is more sensitive 
to the loop contributions of new,
heavy charged particles than  the \ggg\ itself.


The analysis is based on the CompAZ parametrization \cite{compaz} of
the realistic photon collider luminosity spectra.
Presented results correspond to one year of TESLA photon collider
running at nominal luminosity.
The event generation 
according  to the cross-section formula 
\cite{cros_ww,cros_zz}
was done with PYTHIA~6.152.
The fast simulation program SIMDET version 3.01
was used to model the TESLA detector performance.

\begin{figure}[p]
  \begin{center}
  \epsfig{figure=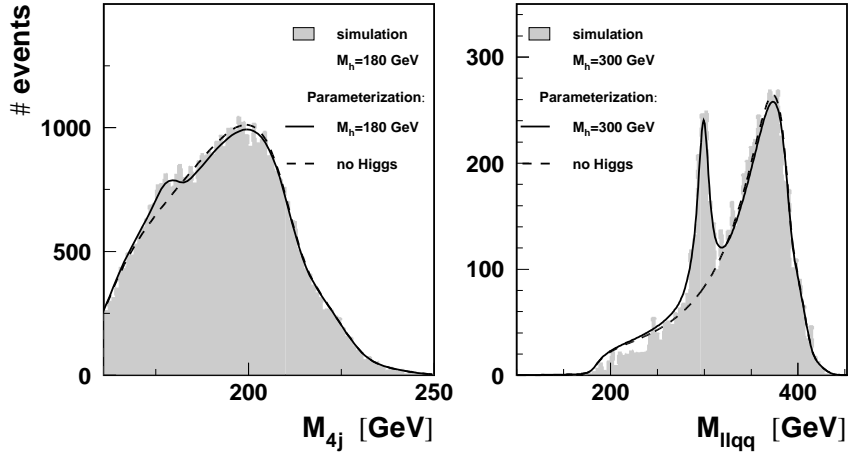,height=6cm,clip=}
  \end{center}
  \vspace{-0.5cm}
  \caption{Distribution of the reconstructed invariant mass
        for $\gamma \gamma \rightarrow W^+ W^-$ events with a 
        SM Higgs-boson mass of 180 GeV and an 
        electron-beam energy of 152.5 GeV (left plot) 
        and
        for $\gamma \gamma \rightarrow \z0  \z0 $ events, with a 
        SM Higgs-boson mass of 300 GeV and an 
        electron-beam energy of 250 GeV (right plot). 
        The distribution expected without the Higgs contribution is
        also shown (dashed line).
          }
  \label{fig:convol}
\end{figure}

\begin{figure}[p]
  \begin{center}
  \epsfig{figure=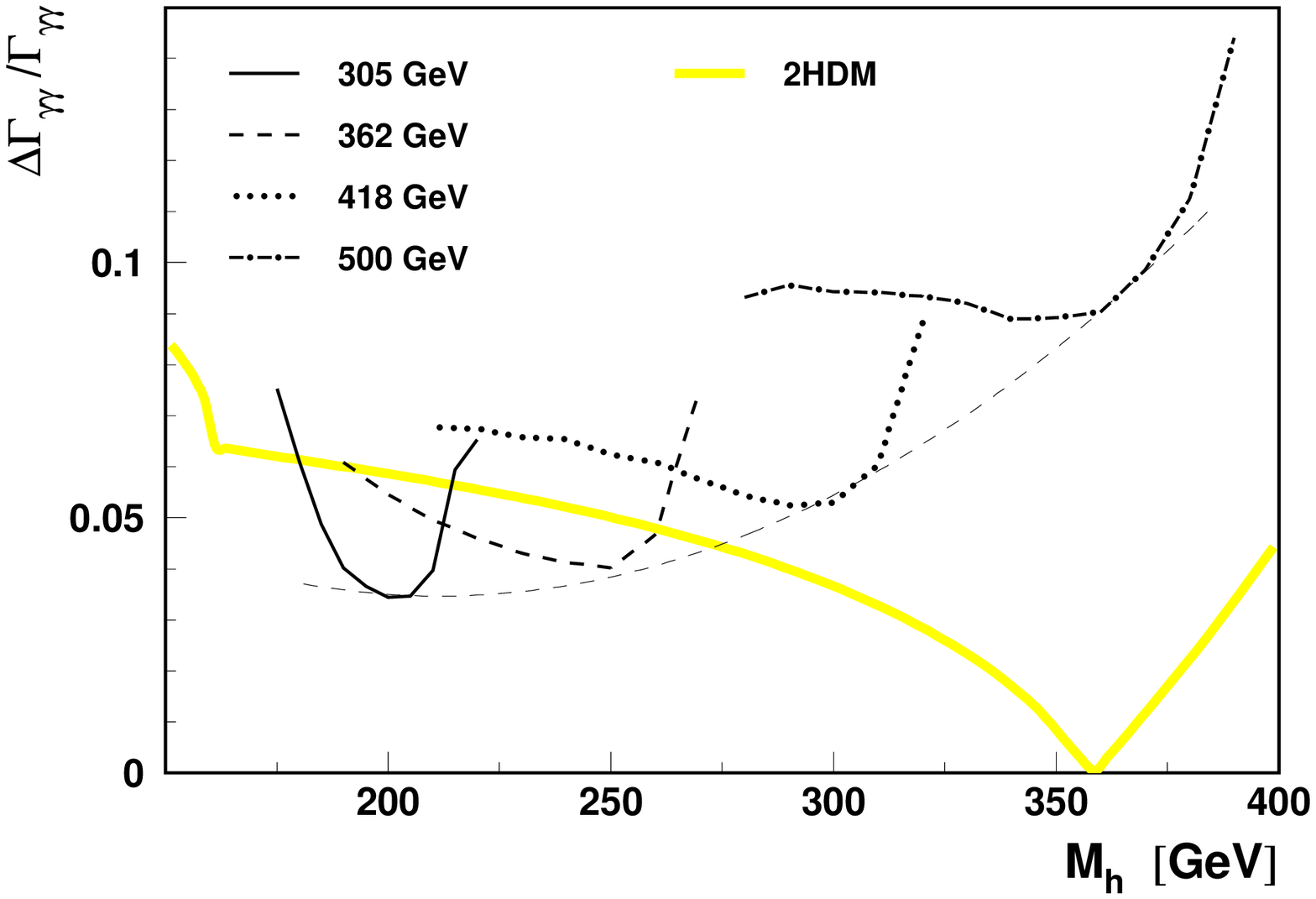,width=6.7cm,clip=}
  \epsfig{figure=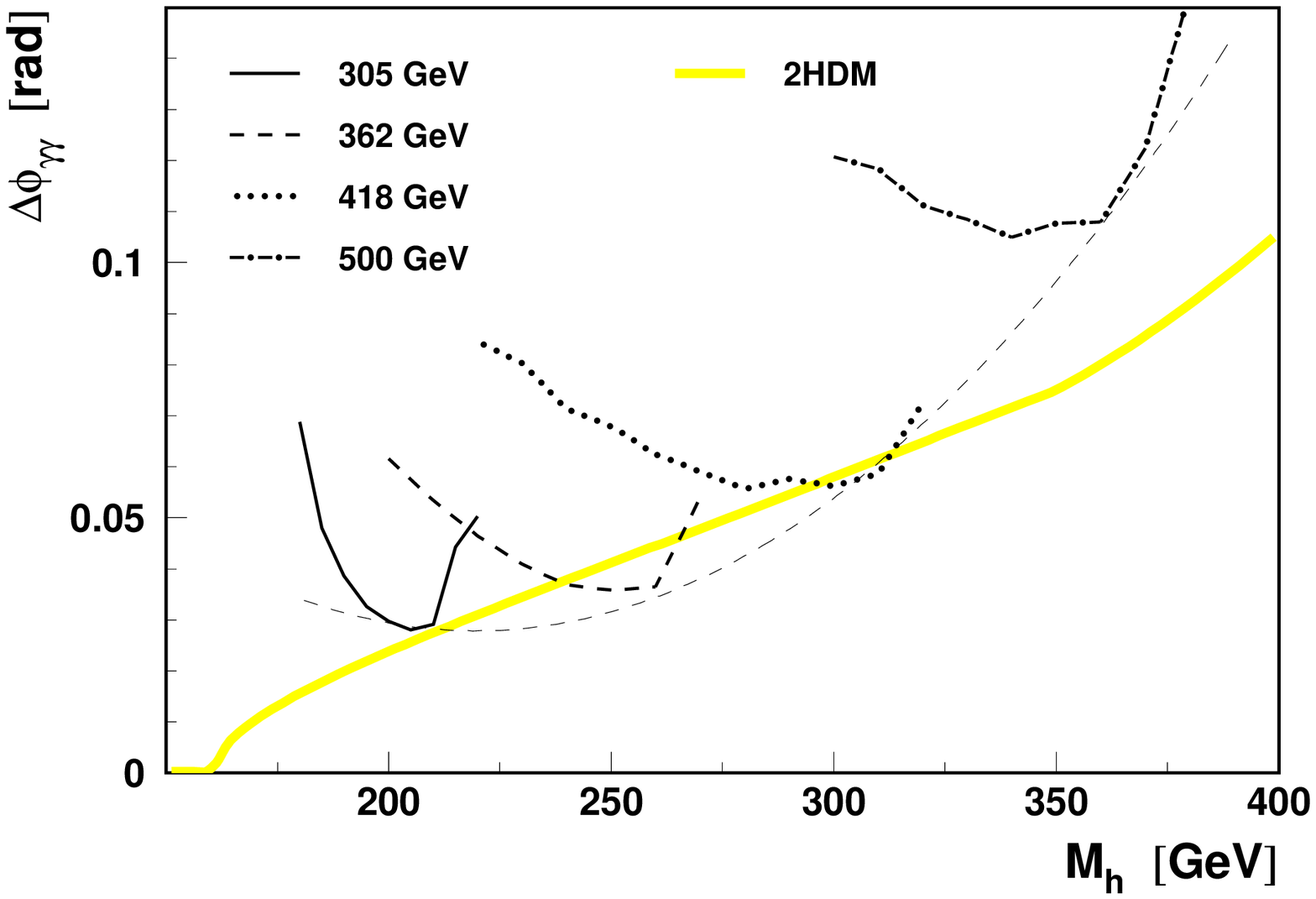,width=6.7cm,clip=}
  \vspace{-0.5cm}
  \end{center}
  \caption{Average statistical error in the 
           determination of the  Higgs-boson width \ggg\ (left plot)
           and phase  \pgg\ (right plot), expected
           after one year of photon collider running, from the simultaneous 
          fit to the observed $W^+ W^-$ and $ZZ$ mass spectra,
          as a function  of the Higgs-boson mass $M_h$.
          The yellow (tick light) band shows the size of 
          the deviations expected in the SM-like  2HDM~(II)~\cite{2HDM}
          with an additional contribution due to the charged
          Higgs-boson of  mass $M_{H^+}=800$ GeV. 
          }
  \label{fig:final_5}
\end{figure}

%
The distribution 
of the reconstructed invariant mass for 
$\gamma \gamma \rightarrow W^+ W^-$ 
 and $\gamma \gamma \rightarrow \z0  \z0 $ events
is shown in  Fig.~\ref{fig:convol}.
 Results coming from the full event simulation  based  on PYTHIA
 and SIMDET  are  compared with the distribution
 obtained  by the numerical convolution of the cross-section 
 formula with the CompAZ spectra and parametrized detector resolution.
%
%
%
Based on the parametric description of the expected mass distributions, 
a number of  experiments were simulated, each corresponding to one year
of TESLA photon collider running at the nominal luminosity.
The ``theoretical'' distributions were then fitted, simultaneously to
the observed $W^+ W^-$ and $\z0 \z0$ mass spectra, with the 
width \ggg\ and phase \pgg\ considered as the only free parameters.
Results of the fits performed for different Higgs-boson masses
and at different electron-beam energies are shown in Fig.~\ref{fig:final_5}.
They indicate that with a proper choice of the beam energy, the $\gamma \gamma$
partial width can be measured with an accuracy of 3 to 8\%, while
the  phase of the amplitude with an accuracy between 30 and 100~mrad,
see Fig. \ref{fig:final_5}.
The \pgg\ measurement opens a new window to  the precise
determination of the Higgs-boson couplings
and to  searches of ``new physics''.

%
%
%

%
%

\end{document}